\documentclass[12pt]{article}

\usepackage{amssymb,amsfonts,amsmath}
\usepackage{graphicx} 
\usepackage{indentfirst}
\usepackage{bbm}

\parskip=\medskipamount

\newcommand {\cH}{{\cal H}}

\newcommand {\cL}{{\cal L}}
\newcommand {\cM}{{\cal M}}
\newcommand {\cN}{{\cal N}}

\newcommand {\cR}{{\cal R}}

\def\a{\alpha}

\def\b{\beta}

\def\d{\delta}
\def\e{\epsilon}

\def\g{\gamma}
\def\G{\Gamma}

\def\k{\kappa}
\def\l{\lambda}

\def\q{\theta}

\def\s{\sigma}

\def\z{\zeta}

\def\F{\Phi}
\def\J{\Psi}
\def\L{\Lambda}
\def\O{\Omega}

\def\S{\Sigma}
\def\U{\Upsilon}

\newcommand{\ad}{{\dot{\alpha}}}                           
\newcommand{\ve}{\varepsilon}                            

\newcommand{\pa}{\partial}                           
\newcommand{\hf}{\frac12}

\newcommand{\sect}[1]{\setcounter{equation}{0}\section{#1}}

\newcommand{\be}{\begin{equation}}
\newcommand{\ee}{\end{equation}}
\newcommand{\bea}{\begin{eqnarray}}
\newcommand{\eea}{\end{eqnarray}}
\newcommand{\non}{\nonumber}

\def\dt#1{{\buildrel {\hbox{\LARGE .}} \over {#1}}}    

\def\double #1{#1{\hbox{\kern-2pt $#1$}}}

\begin{document}

\begin{titlepage}

\begin{flushright}
HIP-2007-46/TH\\
UUITP-16/07\\
NORDITA-2007-27\\
arXiv:0709.2633 [hep-th]
\end{flushright}
\vspace{5mm}

\begin{center}
{\Large \bf  Polar supermultiplets, Hermitian symmetric spaces}\\
{\Large \bf  and hyperk\"ahler metrics}
\end{center}

\begin{center}

{\large  
Masato Arai${},^{a,}$\footnote{masato.arai@helsinki.fi}
Sergei M. Kuzenko${}^{b,}$\footnote{{kuzenko@cyllene.uwa.edu.au}}
and Ulf Lindstr\"om${}^{c,}$\footnote{ulf.lindstrom@teorfys.uu.se} 
} \\
\vspace{5mm}

\footnotesize{
${}^{a}${\it High Energy Physics Division, Department of Physical Sciences\\
University of Helsinki and Helsinki Institute of Physics\\
P.O. Box 64, FIN-00014, Finland
}}  
\\
\vspace{2mm}

\footnotesize{
${}^{b}${\it School of Physics M013, The University of Western Australia\\
35 Stirling Highway, Crawley W.A. 6009, Australia}}  
\\
\vspace{2mm}

\footnotesize{
${}^{c}${\it Department of Theoretical Physics,
Uppsala University \\ 
Box 803, SE-751 08 Uppsala, Sweden\\

 and\\
 
 NORDITA, Roslagstullsbacken 23,\\
 SE-10691 Stockholm, Sweden\\
 
 and\\
 
 Helsinki Institute of Physics\\
P.O. Box 64, FIN-00014, Finland
}}
\\

\vspace{2mm}

\end{center}
\vspace{5mm}

\begin{abstract}
\baselineskip=14pt
\noindent
We address the construction of four-dimensional ${\cal{N}}=2$ 
supersymmetric nonlinear sigma models on tangent bundles of {\it arbitrary} Hermitian symmetric
spaces starting from projective superspace. Using a systematic way of solving
the (infinite number of) auxiliary field equations along with the requirement of 
supersymmetry, we are able to derive a closed form for the Lagrangian on the tangent
bundle and to dualize it to give the hyperk\"ahler potential on the cotangent bundle.
As an application, the case of the exceptional symmetric space $E_6/SO(10) \times U(1)$ 
is explicitly worked out for the first time.
\end{abstract}
\vspace{1cm}

\vfill
\end{titlepage}

\newpage
\setcounter{page}{1}
\renewcommand{\thefootnote}{\arabic{footnote}}
\setcounter{footnote}{0}

\tableofcontents{}
\vspace{1cm}
\bigskip\hrule

\sect{Introduction}

The intimate relation between the number of supersymmetries and 
the target space geometry 
for supersymmetric sigma model \cite{Zumino} has been fruitfully
exploited over the years. 
Here we are interested in the four-dimensional ${\cal{N}}=2$  models whose target space is 
hyperk\"ahler \cite{Alvarez-Gaume:1981hm}. 

There are two methods for constructing new models from old ones; the Legendre transform and 
the hyperk\"ahler reduction \cite{Lindstrom:1983rt,Hitchin:1986ea}, 
both of which have been reformulated in the
manifest ${\cal{N}}=2$ supersymmetric setting of projective superspace. 

Projective superspace extends superspace at each point by an additional bosonic coordinate
 $\zeta$ which is a projective coordinate on $\mathbb{C}P^1$; 
actions are written using contour 
integrals over $\zeta$, and reality conditions are imposed using complex 
conjugation of $\zeta$
composed with the antipodal map \cite{Karlhede:1984vr,LR1,LR2,{G-RLRWvU}}. 
 
In a recent paper \cite{AKL}, 
we constructed, building in part on earlier work  \cite{GK1,GK2},  
$\cN=2$ supersymmetric 
 models on the tangent bundles of a large number of the 
 Hermitian symmetric spaces 
 as well as, using the generalized Legendre transform \cite {LR1}, 
 the hyperk\"ahler metrics
 on the corresponding cotangent bundles. 
Our approach rested on finding solutions to the ${\cal{N}}=2$ projective superspace 
 auxiliary field equations in K\"ahler normal  coordinates at a point and then extending the solutions using cleverly chosen
 coset representatives. Although this method is perfectly viable, it becomes very cumbersome 
 when more complicated spaces
 involving the exceptional groups are considered. 
 ${}$For this reason we have changed the perspective 
 in this paper.  Our discussion is based 
 on the solution to the auxiliary field equations originally described in \cite{GK1,GK2}. 
 Starting from this solution
 and the duly modified second supersymmetry transformation allows us to completely determine the tangent-bundle action.
 We also describe how to find the dual cotangent-bundle action. 
 
As illustrations of our method, we rederive some of the results in
 \cite{AKL}. 
As a new application, we present a model on the tangent bundle of
 $E_{6}/SO(10)\times U(1)$ as well as the hyperk\"ahler potential 
 on the corresponding cotangent bundle.
 
The organization of the paper is as follows. 
In section two we describe
 the background material on $\cN=2$ sigma models
 formulated using  projective superspace. 
 Our general construction is presented in section three. 
 Section four contains the 
 application to $E_{6}/SO(10)\times U(1)$, 
and in section five we give an alternative description of our Lagrangian, which leads to very direct relations to previous
 results but seems to have a more limited applicability. 
Examples are found in section five and in Appendix A.
Appendix B contains an explicit derivation of a relation used in section four.

\sect{Background material on $\cN=2$ sigma models}

We are interested in a family of 4D $\cN=2$ off-shell 
supersymmetric nonlinear sigma-models that are described in 
ordinary $\cN=1$ 
superspace by the action\footnote{The study of such models 
in this context was initiated in \cite{K,GK1,GK2}. 
They correspond to a subclass of the general hypermultiplet theories
in projective superspace \cite{LR1,LR2}.}
\bea
S[\U, \breve{\U}]  =  
\frac{1}{2\pi {\rm i}} \, \oint \frac{{\rm d}\z}{\z} \,  
 \int {\rm d}^8 z \, 
K \big( \U^I (\z), \breve{\U}^{\bar{J}} (\z)  \big) ~.
\label{nact} 
\eea
The action is formulated in terms of 
the so-called {\it polar} multiplet \cite{LR1,LR2} (see also \cite{G-RLRWvU}), 
one of the most interesting $\cN=2$  multiplets living in projective superspace.
The polar multiplet is described by an arctic superfield $\U(\z)$ and 
 antarctic superfield
 $\breve{\U} (\z) $ that 
are generated by an infinite set of ordinary $\cN=1$ superfields:
\be
 \U (\z) = \sum_{n=0}^{\infty}  \, \U_n \z^n = 
\F + \S \,\z+ O(\z^2) ~,\qquad
\breve{\U} (\z) = \sum_{n=0}^{\infty}  \, {\bar
\U}_n
 (-\z)^{-n}~.
\label{exp}
\ee
Here $\F$ is chiral, $\S$  complex linear, 
\be
{\bar D}_{\dt{\a}} \F =0~, \qquad \qquad {\bar D}^2 \S = 0 ~,
\label{chiral+linear}
\ee
and the remaining component superfields are unconstrained complex 
superfields.  
The above  theory
 occurs as a minimal $\cN=2$ extension of the
general four-dimensional $\cN=1$ supersymmetric nonlinear sigma model \cite{Zumino}
\be
S[\F, \bar \F] =  \int {\rm d}^8 z \, K(\Phi^{I},
 {\bar \Phi}{}^{\bar{J}})  ~,
\label{nact4}
\ee
with $K$ the  K\"ahler potential of a K\"ahler manifold $\cM$.

The extended supersymmetric  sigma model  (\ref{nact}) 
inherits  all the geometric features of
its $\cN=1$ predecessor (\ref{nact4}). 
The K\"ahler invariance of the latter,
\be
K(\F, \bar \F) \quad \longrightarrow \quad K(\F, \bar \F)~ +~
\L(\F)+  {\bar \L} (\bar \F) 
\label{kahl}
\ee
turns into 
\be
K(\U, \breve{\U})  \quad \longrightarrow \quad K(\U, \breve{\U}) ~+~
\L(\U) \,+\, {\bar \L} (\breve{\U} ) 
\label{kahl2}
\ee
for the model (\ref{nact}). 
A holomorphic reparametrization of the K\"ahler manifold,
\be
 \F^I  \quad  \longrightarrow   \quad f^I \big( \F \big) ~,
\ee
has the following
counterpart
\be
\U^I (\z) \quad  \longrightarrow  \quad f^I \big (\U(\z) \big)
\label{kahl3}
\ee
in the $\cN=2$ case. Therefore, the physical
superfields of the 
${\cal N}=2$ theory
\be
 \U^I (\z)\Big|_{\z=0} ~=~ \F^I ~,\qquad  \quad \frac{ {\rm d} \U^I (\z) 
}{ {\rm d} \z} \Big|_{\z=0} ~=~ \S^I ~,
\label{kahl4} 
\ee
should be regarded, respectively, as  coordinates of a point in the K\" ahler
manifold and a tangent vector at  the same point. 
Thus the variables $(\F^I, \S^J)$ parametrize the tangent 
bundle $T\cM$ of the K\"ahler manifold $\cM$ \cite{K}. 

To describe the theory in terms of 
the physical superfields $\F$ and $\S$ only, 
all the auxiliary 
superfields have to be eliminated  with the aid of the 
corresponding algebraic equations of motion
\bea
\oint \frac{{\rm d} \z}{\z} \,\z^n \, \frac{\pa K(\U, \breve{\U} 
) }{\pa \U^I} ~ = ~ \oint \frac{{\rm d} \z}{\z} \,\z^{-n} \, \frac{\pa 
K(\U, \breve{\U} ) } {\pa \breve{\U}^{\bar I} } ~ = ~
0 ~, \qquad n \geq 2 ~ .               
\label{asfem}
\eea
Let $\U_*(\z) \equiv \U_*( \z; \F, {\bar \F}, \S, \bar \S )$ 
denote a unique solution subject to the initial conditions
\bea
\U_* (0)  = \F ~,\qquad  \quad \dt{\U}_* (0) 
 = \S ~.
\label{geo3} 
\eea

For a general K\"ahler manifold $\cM$, 
the auxiliary superfields $\U_2, \U_3, \dots$, and their
conjugates,  can be eliminated  only perturbatively. 
Their elimination  can be carried out
using the ansatz \cite{KL}
\bea
\U^I_n = 
\sum_{p=0}^{\infty} 
G^I{}_{J_1 \dots J_{n+p} \, \bar{L}_1 \dots  \bar{L}_p} (\F, {\bar \F})\,
\S^{J_1} \dots \S^{J_{n+p}} \,
{\bar \S}^{ {\bar L}_1 } \dots {\bar \S}^{ {\bar L}_p }~, 
\qquad n\geq 2~.
\eea
Upon elimination of the auxiliary superfields,
the action 
(\ref{nact}) takes the form \cite{GK1,GK2}
\bea
S_{{\rm tb}}[\F,  \S]  
&=& \int {\rm d}^8 z \, \Big\{\,
K \big( \F, \bar{\F} \big)+  
\sum_{n=1}^{\infty} \cL_{I_1 \cdots I_n {\bar J}_1 \cdots {\bar 
J}_n }  \big( \F, \bar{\F} \big) \S^{I_1} \dots \S^{I_n} 
{\bar \S}^{ {\bar J}_1 } \dots {\bar \S}^{ {\bar J}_n }~\Big\}~~~~ \non \\
&\equiv & \int {\rm d}^8 z \, \Big\{\,
K \big( \F, \bar{\F} \big)+ \sum_{n=1}^{\infty} \cL^{(n)}
\big(\F, \bar \F, \S , \bar \S \big)
\Big\}~,
 \label{act-tab}
\eea
where $\cL_{I {\bar J} }=  - g_{I \bar{J}} \big( \F, \bar{\F}  \big) $ 
and the tensors $\cL_{I_1 \cdots I_n {\bar J}_1 \cdots {\bar 
J}_n }$ for  $n>1$ 
are functions of the Riemann curvature $R_{I {\bar 
J} K {\bar L}} \big( \F, \bar{\F} \big) $ and its covariant 
derivatives.  Each term in the action contains equal powers
of $\S$ and $\bar \S$, since the original model (\ref{nact}) 
is invariant under rigid U(1)  transformations \cite{GK1}
\be
\U(\zeta) ~~ \mapsto ~~ \U({\rm e}^{{\rm i} \a} \zeta) 
\quad \Longleftrightarrow \quad 
\U_n(z) ~~ \mapsto ~~ {\rm e}^{{\rm i} n \a} \U_n(z) ~.
\label{rfiber}
\ee

The complex linear tangent variables
$\S$'s in (\ref{act-tab}) can be dualized 
into chiral one-forms, in accordance with the generalized Legendre 
transform \cite{LR1}. The target space for the model thus obtained 
is (an open domain of the zero section) of the cotangent bundle of 
the K\"ahler manifold  $\cM$ \cite{GK1}.

\sect{General construction}

In what follows, we restrict our consideration to the case 
when $\cM$ is a Hermitian symmetric space, hence
\be
\nabla_L  R_{I_1 {\bar  J}_1 I_2 {\bar J}_2}
= {\bar \nabla}_{\bar L} R_{I_1 {\bar  J}_1 I_2 {\bar J}_2} =0~.
\label{covar-const}
\ee
Then, the algebraic equations of motion
(\ref{asfem}) are known to be equivalent  
to the holomorphic geodesic equation
(with complex evolution parameter) \cite{GK1,GK2}
\bea
\frac{ {\rm d}^2 \U^I_* (\z) }{ {\rm d} \z^2 } + 
\G^I_{~JK} \Big( \U_* (\z), \bar{\F} \Big)\,
\frac{ {\rm d} \U^J_* (\z) }{ {\rm d} \z } \,
\frac{ {\rm d} \U^K_* (\z) }{ {\rm d} \z }  =0 ~,
\eea
under the same initial conditions (\ref{geo3}).
Here $\G^I_{~JK} 
( \F , \bar{\F} )$ are the Christoffel symbols for the  
K\"ahler metric $g_{I \bar J} ( \F , \bar{\F} ) = \pa_I 
\pa_ {\bar J}K ( \F , \bar{\F} )$.
In particular, we have 
\bea
\U^I_2 = -\hf \G^I_{~JK} \big( \F, \bar{\F} \big) \, \S^J\S^K~.
\label{U2}
\eea

According to the principles of projective superspace \cite{LR1,LR2}, 
the action  (\ref{nact}) 
is invariant under $\cN=2$ supersymmetry transformations
\be
\d \U (\z)= {\rm i} \left(\ve^\a_i Q^i_\a +
{\bar \ve}^i_\ad {\bar Q}^\ad_i \right)  \U(\z)
\label{SUSY1}
\ee
when $\U (\z)$ is viewed as a $\cN=2$ superfield.
However, since the action is given in $\cN=1$ superspace,
it is only the $\cN=1$ supersymmetry which is manifestly 
realized. The second hidden supersymmetry can be shown to act on the 
physical superfields $\F$ and $\S$ as follows (see, e.g., \cite{G-RLRWvU}):
\bea
\d \F = {\bar \ve}_{\dt{\a}} {\bar D}^{\dt{\a}} \S~, \qquad 
\d \S = -\ve^\a D_\a \F +   {\bar \ve}_{\dt{\a}} {\bar D}^{\dt{\a}} \U_2~.
\eea 
Upon elimination of the auxiliary superfields, 
the action (\ref{act-tab}), which is  associated with the Hermitian symmetric space 
$\cM$, is invariant under 
\bea
\d \F^I = {\bar \ve}_{\dt{\a}} {\bar D}^{\dt{\a}} \S^I~, \qquad 
\d \S^I = -\ve^\a D_\a \F^I -\hf   {\bar \ve}_{\dt{\a}} {\bar D}^{\dt{\a}} 
\Big\{ \G^I_{~JK} \big( \F, \bar{\F} \big) \, \S^J\S^K \Big\} ~.
\label{SUSY2}
\eea 
It turns out that the requirement of invariance under these  transformations
allows one to uniquely determine, by making use of (\ref{covar-const}),
 the tangent-bundle action (\ref{act-tab}). 
One finds 
\bea
\cL^{(1)} &=& - g_{I \bar{J}} \big( \F, \bar{\F}  \big) \S^I {\bar \S}^{\bar{J}} ~, \non \\
\cL^{(n+1)} &\equiv& \cL_{I_1 \cdots I_{n+1} {\bar J}_1 \cdots {\bar 
J}_{n+1} } 
 \S^{I_1} \dots \S^{I_{n+1}} 
{\bar \S}^{ {\bar J}_1 } \dots {\bar \S}^{ {\bar J}_{n+1} } 
\label{recurrence} \\
&=& -\frac{n}{2(n+1)} \,
\cL_{I_1 \cdots I_{n-1} L{\bar J}_1 \cdots {\bar J}_{n} }\,
\S^{I_{n+1}} {\bar \S}^{{\bar J}_{n+1} } \,R_{I_{n+1} {\bar J}_{n+1} I_n }{}^L
\S^{I_1} \dots \S^{I_{n}} 
{\bar \S}^{ {\bar J}_1 } \dots {\bar \S}^{ {\bar J}_{n} } ~.~~~~~~
\non
\eea

It is useful to introduce (conjugate to each other) first-order differential operators 
\bea
\cR_{\S,{\bar \S}} &=& -\hf \S^K {\bar \S}^{ {\bar L} } \,
R_{K {\bar L} I }{}^J   \big( \F, \bar{\F}  \big)\,\S^I \frac{\pa}{\pa \S^J}~, \non \\
{\bar \cR}_{\S,{\bar \S}} &=& \hf  \S^K  {\bar \S}^{ {\bar L} } \,
R_{K {\bar L} {\bar I} }{}^{\bar J} \big( \F, \bar{\F}  \big)\,
{\bar \S}^{\bar I} \frac{\pa}{\pa {\bar \S}^{\bar J}}
=-\hf  \S^K  {\bar \S}^{ {\bar L} } \,
R_{K {\bar L}}{}^{\bar J}{}_ {\bar I}  \big( \F, \bar{\F}  \big)\,
{\bar \S}^{\bar I} \frac{\pa}{\pa {\bar \S}^{\bar J}} \label{fd}
~.
\eea
Since the metric and the  curvature tensor are covariantly constant, 
we have 
\bea
[\nabla_K , {\bar \nabla}_{\bar L} ] 
\cL_{I_1 \cdots I_{n} {\bar J}_1 \cdots {\bar J}_{n} } =0~,
\eea
and hence 
\bea
\cR_{\S,{\bar \S}}\,  \cL^{(n)} = {\bar \cR}_{\S,{\bar \S}} \,\cL^{(n)}~.
\eea
Now, the second relation in (\ref{recurrence}) can be rewritten as follows:
\bea
\cL^{(n+1)} =  \frac{1}{n+1} \, \cR_{\S,{\bar \S}} \, \cL^{(n)}~.
\eea
This leads to 
\bea
\cL \big(\F, \bar \F, \S , \bar \S \big)
&=&\sum_{n=1}^{\infty} \cL_{I_1 \cdots I_n {\bar J}_1 \cdots {\bar 
J}_n }  \big( \F, \bar{\F} \big) \S^{I_1} \dots \S^{I_n} 
{\bar \S}^{ {\bar J}_1 } \dots {\bar \S}^{ {\bar J}_n }
=\sum_{n=1}^{\infty} \cL^{(n)} \non \\
&=& - g_{I \bar{J}} 
 {\bar \S}^{\bar{J}} \,
\frac{ {\rm e}^{\cR_{\S,{\bar \S}}} -1}{ \cR_{\S,{\bar \S} }}\,
 \S^I 
  ~.
\label{closed}
 \eea
 It is useful to rewrite this Lagrangian using an auxiliary variable $t$:
 \bea
 \cL \big(\F, \bar \F, \S , \bar \S \big)=-\int_{0}^1{\rm d} t\,g_{I \bar{J}} \,
 {\bar \S}^{\bar{J}}{\rm e}^{t\cR_{\S,{\bar \S}}} \S^I ~.
 \label{tact}
 \eea

The relations (\ref{recurrence}) can be shown to be equivalent to the 
first-order differential equation 
\bea
\hf R_{K {\bar J} L }{}^I\, \frac{\pa \cL}{\pa \S^I}\, \S^K \S^L
+ \frac{\pa \cL}{\pa {\bar \S}^{\bar J} }
+g_{I \bar{J}}\, \S^I =0~ \label{fd2}
\eea
which is obeyed by $\cL \big(\F, \bar \F, \S , \bar \S \big)$ given in 
(\ref{closed}).
Indeed, the action  (\ref{act-tab}) varies under (\ref{SUSY2})
as follows:
\bea
\d S_{{\rm tb}}[\F,  \S]  
&=& \int {\rm d}^8 z \, \Big\{ 
\frac{\pa \cL}{\pa \F^I} -\frac{\pa \cL}{\pa \S^K}\, \G^K_{~IJ} \S^J\Big\}
{\bar \ve}_{\dt{\a}} {\bar D}^{\dt{\a}} \S^I \non \\
&-& \int {\rm d}^8 z \, \Big\{ 
\hf R_{K {\bar J} L }{}^I\, \frac{\pa \cL}{\pa \S^I}\, \S^K \S^L
+ \frac{\pa \cL}{\pa {\bar \S}^{\bar J} }
+g_{I \bar{J}}\, \S^I \Big\} 
{\bar \ve}_{\dt{\a}} {\bar D}^{\dt{\a}} {\bar \F}^{\bar J}~+~{\rm c.c.}
\eea
Here the variation in the first line vanishes, since the curvature is 
covariantly constant.

To construct a dual formulation, consider the first-order action
\bea
S=  \int {\rm d}^8 z \, \Big\{\,
K \big( \F, \bar{\F} \big)+  \cL
\big(\F, \bar \F, \S , \bar \S \big)
+\J_I \,\S^I + {\bar \J}_{\bar I} {\bar \S}^{\bar I} 
\Big\}~,
\label{f-o}
\eea
where the tangent vector $\S^I$ is now  complex unconstrained, 
while the one-form $\Psi$ is chiral, 
${\bar D}_{\dt \a} \J_I =0$.
This action can be shown to be invariant under the 
following supersymmetry transformations:
\bea
\d \F^I &=&\hf {\bar D}^2 \big\{ \overline{\ve \q} \, \S^I\big\} ~, \non \\
\d \S^I &=& -\ve^\a D_\a \F^I -\hf   {\bar \ve}_{\dt{\a}} {\bar D}^{\dt{\a}} 
\Big\{ \G^I_{~JK} \big( \F, \bar{\F} \big) \, \S^J\S^K \Big\} 
-\hf  \overline{\ve \q} \, \G^I_{~JK} \big( \F, \bar{\F})  \, \S^J {\bar D}^2\S^K 
~, \non \\
\d \J_I &=&- \hf {\bar D}^2 \Big\{ \overline{\ve \q} \, K_I \big( \F, \bar{\F})  \Big\}
+\hf {\bar D}^2 \Big\{ \overline{\ve \q} \, \G^K_{~IJ} \big( \F, \bar{\F} \big)\,
\S^J \Big\} \,\J_K
~.
\label{SUSY3}
\eea
Varying $\S$'s and their conjugates 
 in  (\ref{f-o}) using (\ref{tact}) and properties of the curvatures of
 Hermitian symmetric spaces gives 
\bea
\bar\J_{\bar J}&=&g_{I\bar J}\,{\rm e}^{\cR_{\S,{\bar \S}}} \S^I~,\non\\
\J_{I}&=&g_{I\bar J}\,{\rm e}^{\bar\cR_{\S,{\bar \S}}} \bar\S^{\bar J}~.
\label{slns}
\eea
Inverting these relations should lead to  the cotangent-bundle action
\bea
S_{{\rm ctb}}[\F,  \J]  
&=& \int {\rm d}^8 z \, \Big\{\,
K \big( \F, \bar{\F} \big)+    
\cH \big(\F, \bar \F, \J , \bar \J \big)\Big\}~,
\label{act-ctb}
\eea
where 
\bea
\cH \big(\F, \bar \F, \J , \bar \J \big)&=& 
\sum_{n=1}^{\infty} \cH^{I_1 \cdots I_n {\bar J}_1 \cdots {\bar 
J}_n }  \big( \F, \bar{\F} \big) \J_{I_1} \dots \J_{I_n} 
{\bar \J}_{ {\bar J}_1 } \dots {\bar \J}_{ {\bar J}_n } ~,\non \\
\cH^{I {\bar J}} \big( \F, \bar{\F} \big) 
&=& g^{I {\bar J}} \big( \F, \bar{\F} \big)~.
\label{h}
\eea

On general grounds, the cotangent-bundle
action should be  invariant under the supersymmetry transformations
induced from (\ref{SUSY3})
\bea
\d \F^I &=&\hf {\bar D}^2 \big\{ \overline{\ve \q} \, \S^I  \big(\F, \bar \F, \J , \bar \J \big) \big\} ~, \non \\
\d \J_I &=&- \hf {\bar D}^2 \Big\{ \overline{\ve \q} \, K_I \big( \F, \bar{\F})  \Big\}
+\hf {\bar D}^2 \Big\{ \overline{\ve \q} \, \G^K_{~IJ} \big( \F, \bar{\F} \big)\,
\S^J  \big(\F, \bar \F, \J , \bar \J \big) \Big\} \,\J_K
~,
\label{SUSY4}
\eea
with 
\bea
\S^I  \big(\F, \bar \F, \J , \bar \J \big) 
= \frac{\pa}{\pa \J_I} \, \cH  \big(\F, \bar \F, \J , \bar \J \big) ~.
\eea
The requirement of invariance under such transformations 
can be shown to be equivalent to the following nonlinear equation 
on $\cH$:
\be
\S^I \,  g_{I {\bar J}} - \hf \, \S^K\S^L \,  R_{K {\bar J} L}{}^I \,\J_I =
{\bar \J}_{ \bar J} ~. \label{cot-eq}
\ee
This equation also follows directly from (\ref{fd2})  using the definition of the 
$\J$'s, or if one wants, as a consequence of the  superspace Legendre transform.
(It can be explicitly checked that the relation is satisfied for the expressions in 
(\ref{slns}), as it should).

The relation (\ref{cot-eq}) allows us to uniquely
reconstruct  $\cH \big(\F, \bar \F, \J , \bar \J \big)$ 
formally defined in (\ref{h}).

As a simple illustration of the formalism developed,
 in Appendix A we re-derive the model on the tangent bundle of 
$\mathbb{C}P^n$.
The actual power of our method is revealed in next section where
 it is applied to derive a $\cN=2$ supersymmetric sigma model
 on the tangent  bundle of $E_6/SO(10)\times U(1)$.

\sect{The Hermitian symmetric space $E_6/SO(10)\times U(1)$}
The K\"ahler potential for the Hermitian symmetric space 
 $E_6/SO(10)\times U(1)$ was computed by several 
 groups \cite{Achiman:1984ku,yasui,DV,IKK2}
 in different but equivalent forms. 
Here we will use the K\"ahler potential derived 
 in Ref. \cite{IKK2}
 with the aid of the techniques developed in \cite{BKMU}. 
In order to comply with the notation adopted in \cite{IKK2}, 
we will use Greek letters to label indices,  lower indices for
base-space ($\F^I \to \F_\a$) and tangent ($\S^I \to \S_\a$) variables, 
while upper indices will be used for one-forms ($\J_I \to \J^\a$).

Locally, the symmetric space
$E_6/SO(10)\times U(1)$ can be
 described by complex variables $\F_\a$ transforming in 
 the  spinor representation ${\bf 16}$ of $SO(10)$
 and their conjugates.
\begin{eqnarray}
 \F_\a~,\qquad \bar{\F}^\a:=(\F_\a)^*~, 
 \qquad \quad
 \a=1,\dots,16~.
\end{eqnarray}
The K\"ahler potential is
\begin{eqnarray}
 K(\F,\bar{\F})=\ln\left(1+\bar{\F}^{\a}\F_\a+ \frac{1}{ 8}
 (\bar{\F}^{\a}(\sigma_A )_{\a\b}\bar{\F}^\b)(\F_\g(\sigma_A^\dagger)^{\g\d}
\F_\d)\right)~, \quad A=1,\dots 10~~~
\end{eqnarray}
where $(\sigma_A)_{\a\b}= (\sigma_A)_{\b \a}$
 are the $16\times 16$ sigma-matrices which generate, 
 along with their Hermitian-conjugates,  $(\sigma^\dagger_A)^{\a\b}$, 
the  ten-dimensional Dirac matrices in the Weyl representation. 
 The sigma-matrices obey the anti-commutation relations
\begin{eqnarray}
 (\s_A\s_B^\dagger+\s_B\s_A^\dagger)_\a^{~\b}=2\d_{AB}\d_\a^{~\b}~.
\end{eqnarray}
The K\"ahler metric can  be shown to be 
\begin{eqnarray}
g^\a_{~\b}&=&
{\partial^2 K \over \partial\F_\a \partial
  \bar{\F}^\b}
 ={1 \over Z}\Bigg\{
 \delta_\a^{~\b} + \frac{1}{2}
 (\s_A)_{\a\g}\bar{\F}^\g(\s_A^\dagger)^{\b\d}\F_\d \non \\
 &&+{1 \over Z}\Big(-\bar{\F}^\a\F_\b-{1 \over
 4}\bar{\F}^\a(\s_A)_{\b\g}\bar{\F}^\g(\F^{\rm T} \s_A^\dagger\F)
 -{1 \over 4}(\s_A^\dagger)^{\a\d}\F_\d\F_\b(\bar{\F}^{\rm
T}\s_A\bar{\F}) \non \\
&& \qquad -{1 \over
 16}(\s_B^\dagger)^{\a\d}\F_\d(\s_A)_{\b\g}\bar{\F}^\g(\F^{\rm T}
 \s_A^\dagger \F)(\bar{\F}^{\rm T}\s_B\bar{\F})\Big)\Bigg\}~,
\end{eqnarray}
where $Z=1+\bar{\F}^{\rm T}\F+{1 \over 8}(\F^{\rm T}\sigma_A^\dagger\F)
 (\bar{\F}^{\rm T}\sigma_A \bar{\F})$.
Here we have used the fact that $\s_A $ is symmetric.

Let us  calculate the Lagrangian (\ref{closed}) for the case under consideration.
In our notation, the first-order differential operator 
 defined in (\ref{closed}) is  
\begin{eqnarray}
 {\cal R}_{\S,\bar{\S}}=-{1 \over
  2}\S_\a\bar{\S}^\b\S_\g R^{\a~\g}_{~\b~\d}(g^{-1})^\d_{~\e}\,
 {\partial \over \partial \S_\e}~.
\end{eqnarray}
where $(g^{-1})^\b_{~\a}=(g^\a_{~\b})^{-1}$ is the  inverse metric of 
 $g^{\a}_{~\b}$, that is  $g^{\a}_{~\g}(g^{-1})^{\g}_{~\b}=\delta^\a_{~\b}$.
Since we are considering a symmetric space, it is actually sufficient to 
carry out the calculations of our interest at a particular point, say  at $\F=0$. 
The Riemann tensor at $\F=0$ can be shown to be 
\begin{eqnarray}
 R^{\a~\g}_{~\b~\d}{\Big |}_{\F=0}&=&\partial^g\partial_\d g^{\a}_{~\b}
 -(g^{-1})^\l_{~\k}\partial^\k g^\a_{~\b}\partial_\l g^\g_{~\d}{\Big
 |}_{\F=0} \nonumber \\
 &=&-\d_\d^{~\a}\d_\b^{~\g}
 +\frac{1}{ 2}(\s_A)_{\b\d}(\s_A^\dagger)^{\a\g}-\d_\b^{~\a}\d_\d^{~\g}~.
\end{eqnarray}
Now, simple calculations give
\begin{eqnarray}
{\cal R}_{\S,\bar{\S}}\S_\a&=&|\S|^2\S_\a-{1 \over
  4}(\bar{\S}^{\rm T}\s_A)_\a(\S^{\rm T} \s_A^\dagger\S)~,\nonumber \\
({\cal R}_{\S,\bar{\S}})^2\S_\a&=&2|\S|^4\S_\a
 -{1 \over 2}(\bar{\S}^{\rm T}\s_A)_\a|\S|^2(\S^{\rm T} \s_A^\dagger\S)
 -{1 \over 4}\S_\a|\S^{\rm T} \s_A^\dagger  \S|^2~,\nonumber \\
({\cal R}_{\S,\bar{\S}})^3\S_\a&=&6|\S|^6\S_\a
 -{3 \over 2}|\S|^4(\bar{\S}^{\rm T}\s_A)_\a(\S^{\rm T} \s_A^\dagger\S)
 -{3 \over 2}\S_\a|\S|^2|\S^{\rm T} \s_A^\dagger \S|^2 \nonumber \\
 &&+{3 \over 16}(\bar{\S}^{\rm T}\s_A)_\a(\S^{\rm T} \s_A^\dagger\S)|\S^{\rm T} \s_B^\dagger
  \S|^2~,
\end{eqnarray}
where $|\S|^2=\bar{\S}^\a\S_\a$ and 
 $|\S^{\rm T} \s_A^\dagger \S|^2=(\S^{\rm T}  \s_A^\dagger \S)(\bar{\S}^{\rm
 T}\s_A  \bar{\S})$.
Here we have used the following identity
\begin{eqnarray}
 (\s_A^\dagger \F)^\a(\F \s_A^\dagger \F)=0~
\end{eqnarray}
that follows from the Fierz identity
\begin{eqnarray}
 (\epsilon \s_A^\dagger \psi)(\psi \s_A^\dagger \eta)=-{1 \over
  2}(\epsilon \s_A^\dagger \eta)(\psi \s_A^\dagger \psi)~.
\end{eqnarray}
Making use of the above results gives
\begin{eqnarray}
&&{\cal L}(\F=0, \bar{\F}=0,\S,\bar{\S})=-g^\a_{~\b} 
 {\bar \S}^{\b} \,
\frac{ {\rm e}^{\cR_{\S,{\bar \S}}} -1}{ \cR_{\S,{\bar \S} }}\,
 \S_\a{\Bigg |}_{\F=\bar{\F}=0} \nonumber \\
 &&~~~~~= -|\S|^2-{1 \over 2}|\S|^4
 +{1 \over 8}|\S^{\rm T} \s_A^\dagger\S|^2
 -{1 \over 3}|\S|^6 +{1 \over 8}|\S|^2|\S^{\rm T} \s_A^\dagger  \S|^2
 \nonumber \\ 
 &&~~~~~~~~-{1 \over 4}|\S|^8 +{1 \over 8}|\S|^4|\S^{\rm T}  \s_A^\dagger \S|^2
 -{1 \over 128}|\S^{\rm T} \s_A^\dagger \S|^2|\S^{\rm T} \s_B^\dagger
 \S|^2+\cdots ~
 \end{eqnarray}
Looking at the expression obtained   it is tempting to conjecture
\begin{eqnarray} 
 {\cal L}(\F=0, \bar{\F}=0,\S,\bar{\S})
 =\ln\left(1-|\S|^2+{1 \over 8}|\S^{\rm T} \s_A^\dagger \S|^2\right)~.
\end{eqnarray}
The latter relation extends to 
an arbitrary point $\F$ of the base manifold by replacing
\begin{eqnarray}
 |\S|^2~\rightarrow~ g^\a_{~\b}\S_\a\bar{\S}^\b~,\quad
 {1 \over 8}|\S^{\rm T} \s_A^\dagger\S|^2 ~\rightarrow~ 
 {1 \over 2}(g^\a_{~\b}\S_\a\bar{\S}^\b)^2
 +{1 \over 4}R^{\a~\g}_{~\b~\d}\S_\a\bar{\S}^\b\S_\g\bar{\S}^\d~.
\end{eqnarray}
Then one gets
\begin{eqnarray}
{\cal L}(\F,\bar{\F},\S,\bar{\S})
 &=& - g^\a_{~\b} {\bar \S}^{\b} 
 \frac{ {\rm e}^{\cR_{\S,{\bar \S}}} -1}{ \cR_{\S,{\bar \S}
 }}\,\S_\a \nonumber \\
 &=&\ln\left(1-g^\a_{~\b}\S_\a\bar{\S}^\b+{1 \over 2}(g^\a_{~\b}\S_\a\bar{\S}^\b)^2
 +{1 \over
 4}R^{\a~\g}_{~\b~\d}\S_\a\bar{\S}^\b\S_\g\bar{\S}^\d\right)~.
\label{ll}
\end{eqnarray}
This is actually the correct result for ${\cal L}(\F,\bar{\F},\S,\bar{\S})$.
Indeed, one can check that the RHS of (\ref{ll})
 satisfies the master equation (\ref{fd2})
which in the present case reads
\begin{eqnarray}
 {1\over 2}R^{\a~\g}_{~\b~\d}(g^{-1})^\d_{~\e}{\partial {\cal L} \over \partial
  \S_\e}\S_\a \S_\g + {\partial {\cal L} \over \partial
  \bar{\S}^\b}+g^\a_{~\b}\S_\a=0~. 
  \label{fd3}
\end{eqnarray}
In order to prove this claim, it is sufficient to restrict our 
consideration to $\F=0$.
${}$For the first term in the LHS of (\ref{fd3}),
one finds
\begin{eqnarray}
&&{1\over 2}R^{\a~\g}_{~\b~\d}(g^{-1})^\d_{~\e}{\partial {\cal L} \over \partial
  \S_\e}\S_\a \S_\g{\bigg |}_{\F=0}\nonumber \\
&&~~~~~~~~~~~~~
={1 \over Z}\left(2\S_\b|\S|^2
  -{1 \over 2}(\s_A \bar{\S})_\b(\S^{\rm T} \s_A^\dagger \S)
  -{1 \over 4}\S_\b|\S^{\rm T} \s_A^\dagger\S|^2\right)~,
\end{eqnarray}
and this contribution exactly cancels against the other terms in (\ref{fd3}).

Let us dualize the tangent-bundle action. 
${}$For this purpose 
we consider the following first-order action
\begin{eqnarray}
 S=\int {\rm d}^8z{\Big \{}K(\F,\bar{\F})
 &+& \ln\left(1-g^\a_{~\b}U_\a\bar{U}^\b+{1 \over 2}(g^\a_{~\b}U_\a\bar{U}^\b)^2
 +{1 \over 4}R^{\a~\g}_{~\b~\d}U_\a\bar{U}^\b U_\g\bar{U}^\d\right)
 \nonumber \\
 &+&U_\a \J^\a + \bar{U}^\a \bar{\J}_\a{\Big \}}~,
 \label{parent}
\end{eqnarray}
where the tangent variables $U_\a$ are complex unconstrained superfields,
 and
 the one-forms $\J^\a$ are chiral superfields,
 $\bar{D}_{\dot{\a}}\J=0$.
The variables $U$'s and $\bar{U}$'s can be eliminated with the aid of
 their algebraic equations of motion.
This turns the superfield Lagrangian into the hyperk\"ahler potential
\begin{eqnarray}
 &&H(\F,\bar{\F},\J,\bar{\J})=K(\F,\bar{\F})
 -\ln\left(\L+\sqrt{\L+(g^{-1})^\a_{~\b}\J^\b\bar{\J}_\a}\right)\nonumber \\
 &&~~~+\L+\sqrt{\L+(g^{-1})^\a_{~\b}\J^\b\bar{\J}_\a}-{2((g^{-1})^\a_{~\b}\J^\b\bar{\J}_\a)^2
 +\tilde{R}^{\a~\g}_{~\b~\d}\bar{\J}_\a\J^\b\bar{\J}_\g\J^\d
  \over \L+\sqrt{\L+(g^{-1})^\a_{~\b}\J^\b\bar{\J}_\a}}~, 
   \label{cot-E6}
\end{eqnarray}
where
$\tilde{R}^{\a~\g}_{~\b~\d}=(g^{-1})^{\a}_{~\a^\prime}(g^{-1})^{\b^\prime}_{~\b}
 (g^{-1})^{\g}_{~\g^\prime}(g^{-1})^{\d^\prime}_{~\d}
 R^{\a^\prime~\g^\prime}_{~\b^\prime~\d^\prime}$, and 
\begin{eqnarray}
 \L={1 \over 2}+\sqrt{{1 \over 4}+(g^{-1})^\a_{~\b}\J^\b\bar{\J}_\a
    +2((g^{-1})^\a_{~\b}\J^\b\bar{\J}_\a)^2
    +\tilde{R}^{\a~\g}_{~\b~\d}\bar{\J}_\a\J^\b\bar{\J}_\g\J^\d}~.
\end{eqnarray}
The derivation of the above results is given in Appendix 
\ref{hcp-E6}.

Similar to eq. (\ref{fd3}) in the tangent-bundle formulation, one can check that 
 the hyperk\"ahler potential (\ref{cot-E6}) satisfies the equation (\ref{cot-eq}),
 which  in the present case takes the form
\begin{eqnarray}
 \S_\a g^\a_{~\b}-{1 \over 2}\S_\a\S_\g
  R^{\a~\g}_{~\b~\d}(g^{-1})^\d_{~\e}\J^\e
 =\bar{\J}_\b~. 
 \label{cot-fd}
\end{eqnarray}
To prove this, we again set $\F=0$.
Then, the LHS in (\ref{cot-fd}) 
becomes
\begin{eqnarray}
 \S_\b-{1 \over 2}\left(
  -2(\S_\a \J^\a)\S_\b+{1 \over 2}(\s_A \J)_\b(\S^{\rm
  T}\s_A^\dagger \S)
\right)~.
\label{cal4}
\end{eqnarray}
Making here 
use of (\ref{cal2}), 
 we can express $\J$ in terms of $\S$.
Then we have
\begin{eqnarray}
\S_\b-{1 \over 2}\S_\a\S_\g
 R^{\a~\g}_{~\b~\d}(g^{-1})^\d_{~\e}\J^\e{\bigg |}_{\F=0}
 ={1 \over \O}\left(\S_\b-{1 \over 4}(\s_A \bar{\S})_\b
	       (\S^{\rm T}\s_A^\dagger \S)\right)\,,
\end{eqnarray}
 where $\O$ is given in (\ref{cal3}).
Because of (\ref{cal2}),
the expression obtained is
 exactly $\bar{\J}_\b$ at $\F=0$.

\sect{An alternative formulation}
In this section we give a reformulation of the Lagrangian defined by
 (\ref{recurrence}) which more directly
 relates it to our previous results. 
The reformulation requires certain identities to be satisfied for products of curvatures; 
 we have not been able to determine if these identities are 
 for a general Hermitean symmetric space.
We define the operator $\mathbb{R}$ by
\bea
\mathbb{R}:= \frac 1 2 \Sigma^a \bar \Sigma ^{\bar b}R_{a\bar{b}c}^{~~~d}M^c_d
\eea
where $M$ is the generator of the relevant structure group and acts on $\Sigma$
as a transformation of a vector: $[X_b^aM^b_a, \Sigma^c]=X_b^c\Sigma^b$. 
Here $a$ and $\bar a$ are tangent space indices.
Using this we may in certain cases re-write 
 the Lagrangian 
(\ref{closed}) as
\bea
{\cal L}(\F,\bar{\F},\S,\bar{\S})=
-\eta_{a\bar b}\bar\Sigma^{\bar
 b}\ln(1+\mathbb{R})\mathbb{R}^{-1}\Sigma^a \label{altact}
\eea
where $\eta_{a\bar b}$ is the tangent space metric. 
The inverse $\mathbb{R}^{-1}$ is formal at this stage, 
 but in the concrete examples that we want to consider it is always
 possible to make sense of it.
The structure (\ref{altact}) is possible when the curvature satisfies 
\bea
R_{N\bar J_{1}M\bar J_{2}}R_{I_{1}\bar J_{3}I_{2}}^{~~~~~~N}R_{I_{4}\bar J_{4}I_{3}}^{~~~~~~M}
\propto
R_{N\bar J_{1} I_{1}\bar J_{2}}R_{I_{2}\bar J_{3}M}^{~~~~~~N}R_{I_{4}\bar J_{4}I_{3}}^{~~~~~~M}
\eea
when symmetrized in $I_{1}...I_{4}$ and in $\bar J_{1}...\bar J_{4}$,
and similar relations for higher products of curvatures. This is indeed true for the
case of $\mathbb{C}P^n$ discussed in 
 Appendix A. We find that, at the origin,
\bea
\mathbb{R} \mathbb{R}^{-1}\Sigma^a =\Sigma^a
\eea 
if we take
\bea
\mathbb{R}^{-1}=-\frac{r^2}{\Sigma\bar\Sigma}\delta_{b}^{~c}M_{c}^{~b}
\eea
which inserted in (\ref{altact}) leads to the Lagrangian
\bea
{\cal L}(\F,\bar{\F},\S,\bar{\S})=
-\frac{r^2}{\Sigma\bar\Sigma} \bar\Sigma_{a}\ln(1+\mathbb{R})\Sigma^a \label{CeePeeN}
\eea
where all contractions and lowering of indices is done 
 using $\eta_{a\bar b}=\delta_{ab}$ and we have 
\bea
R_{a\bar b c\bar d}= -\frac 1{r^2}\left( \delta_{ab} \delta_{cd}+\delta_{ad} \delta_{bc}\right)~,
\eea
all evaluated at the origin (see   Appendix A for more details). 
Evaluating the expression (\ref{CeePeeN}) and re-expressing the result at an 
 arbitrary point, we recover the standard form of the Lagrangian;
 (\ref{CP1}).

Another case where the appropriate identities are satisfied is 
 for the $SO(n+2)/SO(n)\times SO(2)$-model discussed in Sec. 6 in \cite{AKL}. 
Here the metric at a point is as in the previous example, 
the curvature tensor at the origin is 
\bea
R_{a\bar b c\bar d}= 2\left( -\delta_{ab} \delta_{cd}
 +\delta_{ac} \delta_{bd}-\delta_{ad} \delta_{bc}\right)~,~~~a=1,\dots,n~.
\eea
We may take
\bea
\mathbb{R}^{-1}=-\frac {1}{\Sigma^2 {\bar{\Sigma}}^2}\Sigma_{b}\bar \Sigma^cM_{c}^{~b}
\eea
to yield the following form of the Lagrangian
\bea
{\cal L}(\F,\bar{\F},\S,\bar{\S})=
\frac 1{{\bar\Sigma}^2}\bar\Sigma_{a}\ln(1+\mathbb{R})\bar\Sigma^a~.\label{EX2}
\eea
Evaluating the expression (\ref{EX2}) and re-expressing the result at an 
 arbitrary point, we recover the standard form of the Lagrangian \cite{AKL}.

\vskip 5mm

\noindent
{\bf Acknowledgements:}\\
MA and SMK are grateful to the Department of Theoretical Physics 
at Uppsala University for hospitality.
SMK also acknowledges hospitality, 
at various stages of this project,
of the Albert Einstein Institute at Golm, 
and  the 5th Simons Workshop at Stony Brook.
SMK is supported in part by the Australian Research Council.
UL acknowledges support by EU grant (Superstring theory)  MRTN-2004-512194 
and by VR grant 621-2006-3365.\\

\begin{appendix}
\sect{Example: Complex projective space}

As a simple example, consider the complex projective  space
${\mathbb C}P^n = SU(n+1) / U(n)$ for which we have 
\be 
K (\F, {\bar \F}) = r^2 \ln \left(1 + \frac{1}{r^2}  
\F^L \overline{\F^L} \right)
~,~~~
g_{I {\bar J}} (\F, \bar \F) =  
\frac{ r^2 \d_{I {J}} }{r^2 + \F^L \overline{\F^L} }
- \frac{  r^2   \overline{\F^I} \F^J  }
{(r^2 + \F^L \overline{\F^L})^2 }~,
\label{s2pot}
\ee
where $I,\bar{J}=1,\dots, n$.
It is sufficient to compute the Riemann curvature at $\F=0$
\bea
R_{I_1 {\bar  J}_1 I_2 {\bar J}_2} \Big|_{\F=0} &=& 
K_{I_1 {\bar  J}_1 I_2 {\bar J}_2} \Big|_{\F=0} 
= -\frac{1}{r^2} \Big\{ \d_{I_1 J_1} \d_{I_2 J_2} 
+\d_{I_1 J_2} \d_{I_2 J_1} \Big\}~,
\eea
with all results below corresponding to the choice $\F=0$.
One gets
\bea
\S^{I_1} {\bar \S}^{ {\bar J}_1 } \S^{I_2}\,
R_{I_1 {\bar J}_1 I_2 {\bar J}_2}  
=-\frac{2}{r^2} |\S|^2 \S^{J_2}~, 
\qquad |\S|^2 = \d_{IJ} \S^{I} {\bar \S}^{ {\bar J} } ~,
\eea
and hence
\bea
\cR_{\S,{\bar \S}} = \frac{1}{r^2} |\S|^2\, 
\S^{L} \frac{\pa}{\pa \S^L}~.
\eea
${}$From here 
\bea
\big(\cR_{\S,{\bar \S}} \big)^n \S^I= n!\,
\frac{|\S|^{2n}}{r^{2n}}\, 
\S^{I}
\eea
and hence
\bea
- g_{I \bar{J}} 
 {\bar \S}^{\bar{J}} \,
\frac{ {\rm e}^{\cR_{\S,{\bar \S}}} -1}{ \cR_{\S,{\bar \S} }}\,
 \S^I 
= r^2 \ln \Big(1 - \frac{1}{r^2} \, g_{I {\bar J }} 
(\F, {\bar \F})\; \S^I {\bar \S}^{\bar J} \Big) ~. \label{CP1}
\eea
This agrees with the previous calculations \cite{GK2,AN}.

\sect{Derivation of (\ref{cot-E6})}\label{hcp-E6}
This appendix is devoted to the derivation of the hyperk\"ahler potential
 (\ref{cot-E6}). Since the base manifold is symmetric space, it is
 sufficient to perform the dualization, for the action (\ref{parent}),
 at $\F=0$. Then, the first order Lagrangian 
\begin{eqnarray}
 {\cal L}=\ln\Omega+U_\a \psi^\a +
  \bar{U}^{\a}\bar{\psi}_{\a}~, \qquad 
 \O=1-\bar{U}^{\rm T} U+{1 \over 8}|U^{\rm T} \s_A  U|^2~, 
 \label{cal3}
\end{eqnarray}
leads to the following equations of motion for $\bar{U}$'s and $U$'s: 
\begin{eqnarray}
 {-U_\a+(\s_A \bar{U})_\a (U^{\rm T}\s_A^\dagger U)/4 \over
  \O}=\bar{\psi}_\a~,\quad
 {-\bar{U}^\a+( \s_A^\dagger U)^\a (\bar{U}^{\rm T}\s_A  \bar{U})/4 \over
  \O}=\psi^\a ~, ~~~
 \label{cal2}
\end{eqnarray}
where $\psi$ is a cotangent vector at $\F=0$
(it is  useful to reserve the notation $\J$ for a one-form at a generic point $\F$ of the base manifold).
These equation imply
\begin{eqnarray}
 \bar{\psi}^{\rm T}\s_A^\dagger \bar{\psi}={U^{\rm T}\s_A^\dagger U \over \O}~,
\qquad
 \psi^{\rm T}\s_A  \psi = {\bar{U}^{\rm T}\s_A  \bar{U} \over \O}~,
\end{eqnarray}
and also
\begin{eqnarray}
 {1 \over 4}+\bar{\psi}^{\rm T}\psi+{1 \over 2}|\psi^{\rm T} \s_A  \psi|^2
=\left({1 \over 2}+{\bar{U}^{\rm T}U \over \O}\right)^2~.
 \label{sol-a}
\end{eqnarray}
By construction, the correspondence between the tangent and cotangent
 variables should be such that $U\rightarrow 0 \Leftrightarrow
 \psi\rightarrow 0$. 
This means that we have to choose the ``plus'' solution of
 (\ref{sol-a}), that is
\begin{eqnarray}
 {\bar{U}^{\rm T}U \over \O}=-{1 \over 2}+\sqrt{{1\over 4}+\bar{\psi}^{\rm
  T}\psi+{1\over 2}|\psi^{\rm T} \s_A  \psi|^2}~.
\end{eqnarray}
Now, the results obtained above can be used to express $\O$ via $\psi$
 and its conjugate.
By definition, we have
\begin{eqnarray}
 {1 \over \O}={1 \over \O^2}-{\bar{U}^{\rm T}U \over \O^2}
+{1\over 8}\left|{\psi^{\rm T} \s_A  \psi \over \O}\right|^2~,
\end{eqnarray}
This is equivalent to 
\begin{eqnarray}
 \left({1 \over \O}-{\Lambda \over 2}\right)^2={\Lambda^2 \over 4}-{1
  \over 8}|\psi^{\rm T} \s_A  \psi|^2~, 
  \label{sol-b}
\end{eqnarray}
where
\begin{eqnarray}
 \Lambda={1 \over 2}+\sqrt{{1\over 4}+\bar{\psi}^{\rm
  T}\psi+{1\over 2}|\psi^{\rm T} \s_A  \psi|^2}~.
\end{eqnarray}
Since for $\psi\rightarrow 0$ we should have $\O\rightarrow 1$, it is
 necessary to choose the ``plus'' solution of (\ref{sol-b}), that is
\begin{eqnarray}
 {1 \over \O}={\Lambda \over 2}+\sqrt{{\Lambda^2\over 4}
 -{1\over 8}|\psi^{\rm T} \s_A  \psi|^2}
 ={\Lambda \over 2}+{1 \over 2}\sqrt{\Lambda+\bar{\psi}^{\rm T}\psi}~.
\end{eqnarray}
The above consideration corresponds to the origin, $\F=0$, of the base
 manifold.
To extend these results to an arbitrary point $\F$ of the base
 manifold, we should replace
\begin{eqnarray}
 &\bar{\psi}^{\rm T}\psi ~\rightarrow~
  (g^{-1})^\a_{~\b}\J^\b\bar{\J}_\a~,&\nonumber \\
 & \displaystyle {1\over 8}|\psi^{\rm T} \s_A  \psi|^2 ~\rightarrow ~
{1 \over
  2}((g^{-1})^\a_{~\b}\J^\b\bar{\J}_\a)^2
 +{1 \over 4}\tilde{R}^{\a~\g}_{~\b~\d}\bar{\J}_\a\J^\b\bar{\J}_\g\J^\d~.&
\end{eqnarray}
As a result,  we arrive at (\ref{cot-E6}).
\end{appendix}

\small{

}

\begin{thebibliography}{66}

\bibitem{Zumino}
%
  B.~Zumino,
  ``Supersymmetry and K\"ahler manifolds,''
  Phys.\ Lett.\ B {\bf 87}, 203 (1979).
%
\bibitem{Alvarez-Gaume:1981hm}
  L.~Alvarez-Gaum\'e and D.~Z.~Freedman,
  ``Geometrical structure and ultraviolet finiteness in 
  the supersymmetric sigma model,''
  Commun.\ Math.\ Phys.\  {\bf 80}, 443 (1981).
%
\bibitem{Lindstrom:1983rt}
  U.~Lindstr\"om and M.~Ro\v{c}ek,
  ``Scalar tensor duality and N=1, N=2 nonlinear sigma models,''
  Nucl.\ Phys.\ B {\bf 222}, 285 (1983).
%
\bibitem{Hitchin:1986ea}
  N.~J.~Hitchin, A.~Karlhede, U.~Lindstr\"om and M.~Ro\v cek,
  ``Hyperk\"ahler metrics and supersymmetry,''
  Commun.\ Math.\ Phys.\  {\bf 108}, 535 (1987).
%
\bibitem{Karlhede:1984vr}
  A.~Karlhede, U.~Lindstr\"om and M.~Ro\v cek,
  ``Self-interacting tensor multiplets in N=2 superspace,''
  Phys.\ Lett.\ B {\bf 147}, 297 (1984).
%
\bibitem{LR1}
  U.~Lindstr\"om and M.~Ro\v{c}ek,
  ``New hyperk\"ahler  metrics  and new supermultiplets,''
  Commun.\ Math.\ Phys.\  {\bf 115}, 21 (1988).
%
\bibitem{LR2}
  U.~Lindstr\"om and M.~Ro\v{c}ek,
  ``N=2 super Yang-Mills theory in projective superspace,''
  Commun.\ Math.\ Phys.\  {\bf 128}, 191 (1990).
%
\bibitem{G-RLRWvU}
  F.~Gonzalez-Rey, U.~Lindstr\"om, M.~Ro\v{c}ek, S.~Wiles and R.~von Unge,
  ``Feynman rules in N = 2 projective superspace. I: Massless
  hypermultiplets,''
  Nucl.\ Phys.\  B {\bf 516}, 426 (1998)
  [hep-th/9710250].
%
\bibitem{AKL}
  M.~Arai, S.~M.~Kuzenko and U.~Lindstr\"om,
  ``Hyperk\"ahler sigma models on cotangent bundles of Hermitian symmetric
  spaces using projective superspace,''
  JHEP {\bf 0702}, 100 (2007)
  [hep-th/0612174].
%
\bibitem{GK1}
  S.~J.~Gates Jr. and S.~M.~Kuzenko,
  ``The CNM-hypermultiplet nexus,''
  Nucl.\ Phys.\ B {\bf 543}, 122 (1999) [hep-th/9810137].
%
\bibitem{GK2}
  S.~J.~Gates Jr. and S.~M.~Kuzenko,
  ``4D N = 2 supersymmetric off-shell sigma models on the cotangent  
  bundles of  K\"ahler manifolds,''
  Fortsch.\ Phys.\  {\bf 48}, 115 (2000)
  [hep-th/9903013].

\bibitem{K}
  S.~M.~Kuzenko,
  ``Projective superspace as a double-punctured harmonic superspace,''
  Int.\ J.\ Mod.\ Phys.\ A {\bf 14}, 1737 (1999)
  [hep-th/9806147].

%
\bibitem{KL}
  S.~M.~Kuzenko and W.~D.~Linch,
  ``On five-dimensional superspaces,''
  JHEP {\bf 0602}, 038 (2006)
  [hep-th/0507176].
%
\bibitem{Achiman:1984ku}
  Y.~Achiman, S.~Aoyama and J.~W.~van Holten,
  ``The nonlinear supersymmetric sigma model on E6/SO(10)xU(1),''
  Phys.\ Lett.\  B {\bf 141}, 64 (1984);
  ``Gauged supersymmetric sigma models and E6/SO(10)xU(1),''
  Nucl.\ Phys.\  B {\bf 258}, 179 (1985).
%
\bibitem{yasui}
  Y.~Yasui,
  ``The K\"ahler potential of E6/Spin(10)xSO(2),''
  Prog.\ Theor.\ Phys.\  {\bf 72}, 877 (1984).
%
\bibitem{DV}
  F.~Delduc and G.~Valent,
  ``Classical and quantum structure of the compact K\"ahlerian sigma models,''
  Nucl.\ Phys.\ B {\bf 253}, 494 (1985).
%
\bibitem{IKK2}
  K.~Itoh, T.~Kugo and H.~Kunitomo,
  ``Supersymmetric nonlinear Lagrangians of K\"ahlerian coset spaces G/H: G = E6,
  E7 and E8,''
  Prog.\ Theor.\ Phys.\  {\bf 75}, 386 (1986).
%
\bibitem{BKMU}
  M.~Bando, T.~Kuramoto, T.~Maskawa and S.~Uehara,
  ``Nonlinear realization in supersymmetric theories,''
    Prog.\ Theor.\ Phys.\  {\bf 72}, 313 (1984);\\
  ``Nonlinear realization in supersymmetric theories. 2,''
    Prog.\ Theor.\ Phys.\  {\bf 72}, 1207 (1984).
%
\bibitem{AN}
  M.~Arai and M.~Nitta,
  ``Hyper-K\"ahler sigma models on (co)tangent bundles with SO(n) isometry,''
  Nucl.\ Phys.\ B {\bf 745}, 208 (2006) 
  [hep-th/0602277].
\end{thebibliography}
\end{document}